# Optoelectronic and Excitonic Properties of Oligoacenes: Substantial Improvements from Range-Separated Time-Dependent Density Functional Theory

*Bryan M. Wong*[*,a] *and Timothy H. Hsieh*[b]

[a] Materials Chemistry Department, Sandia National Laboratories, Livermore, California 94551

[b] Department of Physics, Massachusetts Institute of Technology, Cambridge, Massachusetts 02139

*Corresponding author. E-mail: bmwong@sandia.gov

The optoelectronic and excitonic properties in a series of linear acenes (naphthalene up to heptacene) are investigated using range-separated methods within time-dependent density functional theory (TDDFT). In these rather simple systems, it is well-known that TDDFT methods using conventional hybrid functionals surprisingly fail in describing the low-lying $L_a$ and $L_b$ valence states, resulting in large, growing errors for the $L_a$ state and an incorrect energetic ordering as a function of molecular size. In this work, we demonstrate that the range-separated formalism largely eliminates both of these errors and also provides a consistent description of excitonic properties in these systems. We further demonstrate that re-optimizing the percentage of Hartree-Fock exchange in conventional hybrids to match wavefunction-based benchmark calculations still yields serious errors, and a full 100% Hartree-Fock range separation is essential for simultaneously describing both of the $L_a$ and $L_b$ transitions. Based on an analysis of electron-hole transition density matrices, we finally show that conventional hybrid functionals overdelocalize excitons and underestimate quasiparticle energy gaps in the acene systems. The results of our present study emphasize the importance of both a range-separated and asymptotically-correct contribution of exchange in TDDFT for investigating optoelectronic and excitonic properties, even for these simple valence excitations.

**1. Introduction**

Conjugated organic structures have attracted significant recent attention due to their potential applications in single-molecule transistors and organic photovoltaics. In the quest for smaller and more efficient electronics, organic semiconductors serve as a promising alternative to their silicon counterparts because of their increased electronic efficiency[1-5] and ease of chemical functionalization.[6-10] In this context, oligoacenes which are composed of linearly fused benzene rings (Figure 1) have high application potential since they possess large charge-carrier mobilities and tunable electronic band gaps. Most notably, pentacene is already utilized as an organic field-effect transistor due to its large hole mobility (5.5 $cm^2$/V·s) which exceeds that of amorphous silicon.[11-13] In general, the linear acenes are especially important since they form the basic fundamental units of armchair graphene nanoribbons which continue to garner enormous interest as novel nanoscale materials.[14-19]

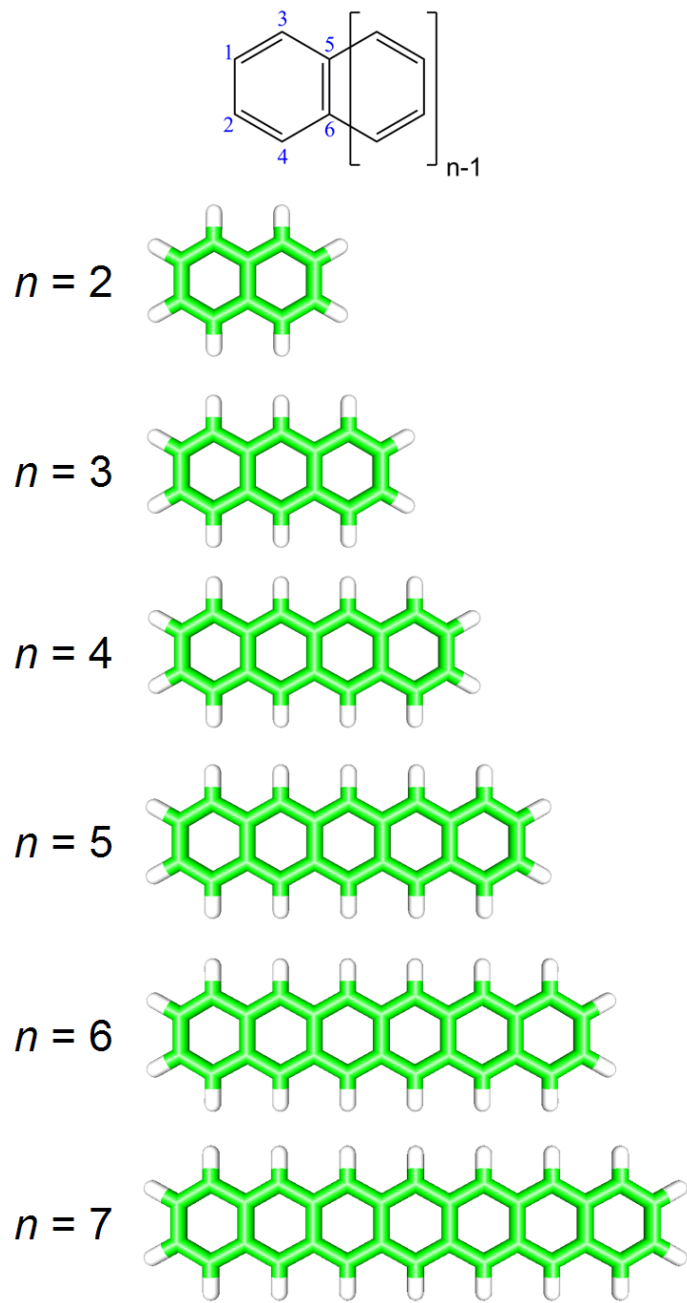

**Figure 1.** Molecular structure and atom labels for the linear acenes. The specific atom numbers depicted in this figure define an ordered basis for generating the transition density matrices discussed in section 3.

In addition to their promising photovoltaic applications, the oligoacenes are also noteworthy as a unique system in which the successes and failures of time-dependent density functional theory (TDDFT) can be assessed and addressed. In 2003, Grimme and Parac noted a dramatic failure (> 0.5 eV error in excitation energies) of TDDFT using standard hybrid functionals for the lowest-lying $\pi \rightarrow \pi^*$ states of the oligoacenes.[20] Their findings were particularly unusual since these types of valence excitations are typically well described (within 0.1 eV) by hybrid TDDFT calculations. While it is well-known that long-range charge-transfer and Rydberg excitations provide a significant challenge for TDDFT,[21-32] these effects are not present in the acene systems since none of the valence excitations possess Rydberg character or involve any long-range charge transfer (both the ground- and excited-state dipole moments are exactly zero by molecular symmetry). As a result, the unexpected failure of TDDFT in these simple valence excitations is most unusual and somewhat surprising.

The present study has two aims. First, we show that certain range-separated functionals,[33-46] which incorporate *both* a position-dependent admixture and an asymptotically correct contribution of Hartree Fock (HF) exchange, yield substantial improvements over conventional hybrids for the various oligoacene excitations. Numerical optimization of parameters in the range-separated and hybrid functionals are carried out to understand their effect on excitation energies and their overall trends. Following the two-dimensional real-space analysis approach of Tretiak et al.,[47-50] we then examine excitonic effects for the various excitations and TDDFT methods. The transition densities and electron difference density maps enable us to understand why conventional hybrids fail and how range-separated functionals accurately reproduce excitation energies and quasiparticle energy gaps for each of the different transitions. We begin by briefly reviewing these two different formalisms and then compare their accuracy in predicting oligoacene excitation properties.

## 2. Theory and Methodology

**2.1 Global Hybrid Functionals.** Recall that DFT is an exact theory in which the only inaccuracies encountered in practice arise from approximations to the (still unknown) exchange-correlation functional. One of the most widely-used DFT schemes for the exchange-correlation energy is Becke's three-parameter B3LYP method[51] which has a relatively simple formulation given by

$$E_{xc}^{\text{global}} = a_0 E_{x,\text{HF}} + (1-a_0) E_{x,\text{Slater}} + a_x \Delta E_{x,\text{Becke88}} + (1-a_c) E_{c,\text{VWN}} + a_c \Delta E_{c,\text{LYP}}. \tag{1}$$

In this expression, $E_{x,\text{HF}}$ is the HF exchange energy based on Kohn-Sham orbitals, $E_{x,\text{Slater}}$ is the uniform electron gas exchange-correlation energy,[52] $\Delta E_{x,\text{Becke88}}$ is Becke's 1998 generalized gradient approximation (GGA) for exchange,[53] $E_{c,\text{VWN}}$ is the Vosko-Wilk-Nusair 1980 correlation functional,[54] and $\Delta E_{c,\text{LYP}}$ is the Lee-Yang-Parr correlation functional.[55] Depending on the choice of the GGA, there are other numerous hybrid functionals in the literature which combine different GGA treatments of exchange and correlation with varying coefficients. In these "global hybrid" functionals, the fraction of nonlocal HF exchange, $a_0$, is held constant in space and fixed to a GGA-specific value (the B3LYP functional, for example, is parameterized with $a_0 = 0.20$).

**2.2 Range-Separated Functionals.** In contrast to conventional hybrids which incorporate a constant fraction of HF exchange, the long-range-corrected[35,37,40,41] (abbreviated as LC or LRC in the literature) formalism mixes HF exchange densities non-uniformly by partitioning the electron repulsion operator as

$$\frac{1}{r_{12}} = \frac{1-\text{erf}(\mu r_{12})}{r_{12}} + \frac{\text{erf}(\mu r_{12})}{r_{12}}. \tag{2}$$

The "erf" term denotes the standard error function, $r_{12} = |\mathbf{r}_1 - \mathbf{r}_2|$ is the interelectronic distance between electrons at coordinates $\mathbf{r}_1$ and $\mathbf{r}_2$, and $\mu$ is the range-separation parameter in units of Bohr$^{-1}$. The first term in Equation 2 is a short-range interaction which decays rapidly on a length scale of $\sim 2/\mu$, and the second term is the long-range part of the Coulomb potential. For a general GGA or hybrid functional, the corresponding exchange-correlation energy according to the LC formalism is

$$E_{xc}^{\text{LC}} = E_{c,\text{DFT}} + (1-a_{\text{HF}}) E_{x,\text{DFT}}^{\text{SR}} + a_{\text{HF}} E_{x,\text{HF}}^{\text{SR}} + E_{x,\text{HF}}^{\text{LR}}, \tag{3}$$

In this expression, $E_{c,\text{DFT}}$ is the original, unmodified DFT correlation contribution, $E_{x,\text{DFT}}^{\text{SR}}$ and $E_{x,\text{HF}}^{\text{SR}}$ are the respective DFT and HF contributions computed with the short-range part of the Coulomb operator (first term in Equation 2), and $E_{x,\text{HF}}^{\text{LR}}$ is the HF exchange contribution evaluated using the long-range part of the Coulomb potential[56] (second term in Equation 2). The $a_{\text{HF}}$ parameter is the coefficient of HF exchange present in the original hybrid functional ($a_{\text{HF}} = 0$ if the original functional is a pure density functional, i.e. BLYP, BOP, or PBE).

It is important to mention at this point that there are also several other range-separation techniques and functionals in the literature, and that the prescription given in equations (2) – (3) is only one of many LC forms. For example, the range-separation technique has been further modified by Handy *et al.*[39,42] with their CAM-B3LYP (Coulomb-attenuating method-B3LYP) methods. Similarly, the Scuseria group has also developed several new range-separated functionals based on a semilocal exchange-hole approach.[43-46] These exchange-hole models have been further refined by the Herbert group to design new functionals which accurately describe both ground- and excited-states.[25,28] In terms of chemical applications, Jacquemin *et al.* have also presented benchmarks for several families of excitations including the electronic spectra of anthroquinone dyes,[57] $n \to \pi^*$ transitions in nitroso and thiocarbonyl dyes,[58] and $\pi \to \pi^*$ excitations in organic chromophores.[59] Very recently, there has also been pioneering work by the Baer and Kronik groups in constructing range-separation functionals tuned entirely from first-principles.[29,30] The key to their success is the choice of a range-separation parameter, $\mu$, which minimizes the difference between the ionization energy (IE) and the negative of the highest-occupied molecular orbital (HOMO) energy, $-E_{\text{HOMO}}$, of the molecule. Since the ionization energy is rigorously equal to $-E_{\text{HOMO}}$ for an "exact functional," the formalism by Baer and Kronik is entirely self-consistent and does not require any experimental input or high-level benchmark calculations.

In all of these various range-separated methods, the key improvement in their accuracy is the smooth separation of DFT and nonlocal HF exchange interactions through the parameter $\mu$. Specifically, the exchange-correlation potentials of conventional density functionals exhibit the wrong asymptotic behavior, but the LC scheme ensures that the exchange potential smoothly recovers the exact $-1/r$ dependence at large interelectronic distances. It is important to point out that the length-scale partitioning in the LC formalism is essential for obtaining accurate TDDFT results. More precisely, a 100% global HF exchange fraction without range separation can corrupt the delicate balance between exchange and correlation contributions, resulting in large errors in excitation energies. For extended charge-transfer processes, the long-range exchange corrections are also particularly vital since these types of excitations are especially sensitive to the asymptotic part of the nonlocal exchange-correlation potential.

**2.3 Computational Details.** For the linear acenes in this work, we compared the performance of global hybrid functionals against range-separated and wavefunction-based calculations. In order to investigate the role of different HF exchange schemes in the various TDDFT methods, we explored the effect of changing the HF exchange fraction, $a_0$, in the global hybrid model and the result of varying the range-separation parameter $\mu$ within the LC formalism. For the parametric study on global hybrids, we kept the same functional form in Becke's three-parameter model (Equation 1) and computed vertical singlet excitation energies as a function of $a_0$ ranging from 0.0 to 1.0 in increments of 0.05. In these calculations, we fixed $a_x = 1 - a_0$ in Equation 1 but kept the correlation contribution with $a_c = 0.81$ unchanged. The $a_x = 1 - a_0$ convention is a common choice used in many hybrid functionals[60-63] such as Becke's B1 convention[61] (in a previous study on large oligothiophenes,[31] we had carried out calculations with $a_x$ fixed to the original 0.72 value recommended by Becke and found that all of the excitation energies were nearly identical compared to the $a_x = 1 - a_0$ convention).

To explore the effect of range-separated exchange on the optoelectronic properties of the acenes, we computed vertical singlet excitation energies as a function of $\mu$ ranging from 0 to 0.90 Bohr$^{-1}$ (in increments of 0.05 Bohr$^{-1}$) while keeping the correlation contribution $E_{c,\text{DFT}}$ in the LC-BLYP functional unchanged. In our study, we utilized several range-separated functionals including CAM-B3LYP, LC-BOP, LC-PBE, LC-ωPBE, and LC-BLYP but found that all of the full-HF-exchange LC functionals gave similar results for the linear acenes. It is very important to note that the original CAM-B3LYP functional is defined[39] with a coulomb-attenuating parameter of $\alpha + \beta = 0.65$ and, therefore, exhibits a – $0.65/r$ dependence for the exchange potential. As a result, the CAM-B3LYP functional is particularly different than the other LC functionals considered in this work since it does not incorporate a full 100%

HF exchange at large interelectronic distance. The very similar results obtained from the other full-exchange LC functionals imply that the excitation energies are not very sensitive to the specific DFT correlation contribution used, and that the systematic error observed previously by Grimme and Parac[20] is largely due to the HF exchange component for the acene systems. In light of these similarities, much of our parametric study focuses on the LC-BLYP results since the other full-HF-exchange LC methods give very similar energies as a function of $\mu$. We should also note that a direct comparison between the LC-BLYP, CAM-B3LYP, and the global hybrid model in Equation 1 allows a very fair and consistent evaluation since all of these methods have similar correlation contributions.

As benchmarks for assessing the quality of the various TDDFT methods, we calculated CC2/cc-pVTZ excitation energies for the linear acenes ranging from $n = 2$ to 7 benzene rings (we stop at $n = 7$ since our CC2/cc-pVTZ calculations indicate a very abrupt and large multi-reference/di-radical character for $n = 8$). We use the CC2 excitation energies as reference values since EOM-CCSD and CASPT2 calculations with the cc-PVTZ basis set were out of reach for larger acenes containing 5 or more benzene rings. Furthermore, we consider the CC2 results as reliable reference values since they accurately reproduce solvent-corrected experimental excitation energies[20] (see Table 1) and are close to CC3 benchmark calculations for the smaller acenes.[64] As an additional check on the quality of the CC2 calculations, we found that none of the acene systems required a multi-reference treatment of electron correlation (D1 diagnostic values were in the 0.04 – 0.06 range), and contributions from single excitations were always greater than 90%.

In order to maintain a consistent comparison across the B3LYP, CAM-B3LYP, LC-BOP, LC-PBE, LC-ωPBE, LC-BLYP, and CC2 levels of theory, identical molecular geometries were used for each of these methods. These reference geometries were optimized at the B3LYP/cc-PVTZ level of theory and are available in the Supporting Information. For all of the TDDFT excitation energies, we used a cc-PVTZ basis set and a high-accuracy Lebedev grid consisting of 96 radial and 302 angular quadrature points. All TDDFT calculations were performed with a locally modified version of GAMESS,[65] and the CC2 calculations were carried out with the TURBOMOLE package.[66]

## 3. Results and Discussion

We focus on two different valence excitations in the linear acenes, commonly labeled in the literature[20,67] as $L_a$ (lowest excited state of $B_{2u}$ symmetry) and $L_b$ ($B_{3u}$ symmetry). The $L_a$ excited state results from a HOMO → LUMO transition with polarization along the molecular short axis, and the $L_b$ state is characterized by a nearly equal mixture of HOMO-1 → LUMO and HOMO → LUMO+1 excitations with a total polarization along the long axis.[68] Using the CC2 excitations as reference values, we performed a total root-mean-square error (RMSE) analysis for all 12 energies (6 $L_a$ and 6 $L_b$ transitions) as a function of $\mu$ and $a_0$. As seen in Figure 2a, the RMSE curve for LC-BLYP has a minimum at $\mu = 0.29$ Bohr$^{-1}$ with a RMS error of 0.10 eV. Perhaps, surprisingly, this RMSE-optimized value of $\mu$ is nearly identical to the 0.31 Bohr$^{-1}$ value recommended for simultaneously describing excitation and fluorescence energies in large oligothiophenes.[31] The RMSE in Figure 2b for the B3LYP-like global hybrid functional has a minimum at $a_0 = 0.50$, with a larger error of 0.20 eV. It is worth noting that our RMSE-minimization with $a_0 = 0.50$ and $a_x = 1 - a_0$ yields a functional very similar to the BHHLYP functional (originally defined with $a_c = 0$) with the exception that our choice has an extra correlation contribution due to the $\Delta E_{c,\text{LYP}}$ term in Equation 1. We denote this re-optimized hybrid functional with $a_0 = 0.50$ as B3LYP$_{\text{opt}}$ in the remainder of this work. Unless otherwise noted, all further LC-TDDFT calculations indicate a range-separation parameter of $\mu = 0.29$ Bohr$^{-1}$.

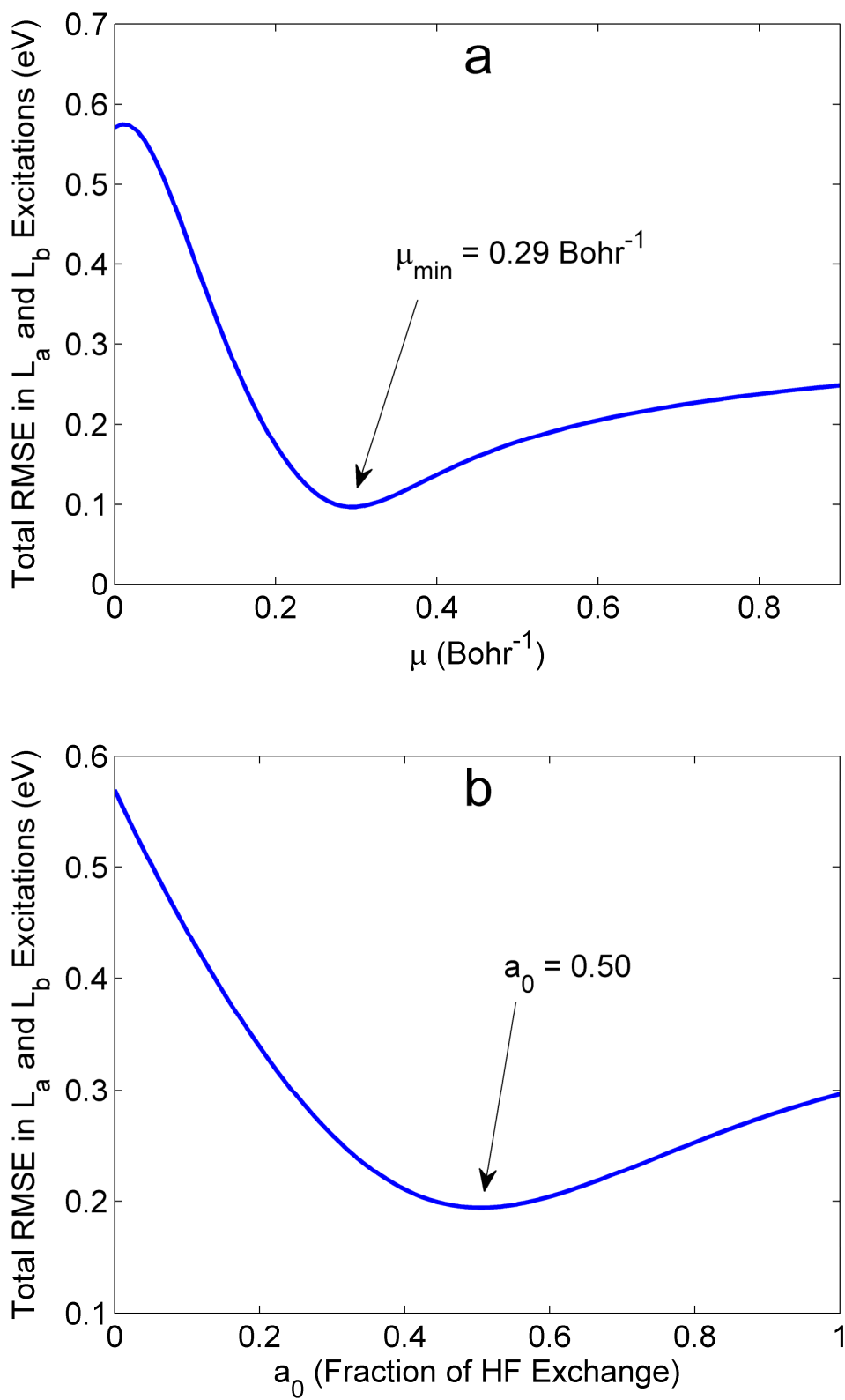

**Figure 2.** Total root-mean-square errors (RMSE) as a function of (a) the range-separation parameter $\mu$ in the LC-BLYP functional and (b) the HF exchange fraction $a_0$ in a B3LYP-like hybrid functional. Figure 3a shows the RMSE curve having a minimum at $\mu = 0.29$ Bohr$^{-1}$, and Figure 3b shows the RMSE curve having a minimum at $a_0 = 0.50$.

Table 1 compares the $L_a$ and $L_b$ excited-state energies between B3LYP, B3LYP$_{opt}$, CAM-B3LYP, LC-BOP, LC-PBE, LC-ωPBE, LC-BLYP, CC2, and Figures 3a – 3b depict in more detail the general trends in transition energies (expressed in wavelength units) between the various TDDFT and CC2 results. It is most important to note in these figures that the energetic ordering of the two electronic states is different, depending on the size of the acene. Specifically, both CC2 and experimental studies[69] indicate a curve crossing between the $L_a$ and $L_b$ states occurs slightly before $n = 3$ benzene rings (anthracene). For all of the other larger acenes, the $L_a$ state lies energetically below the $L_b$ state. Examining Table 1 and Figure 3b, we find that the full-exchange LC-TDDFT calculations are unique in that they show excellent agreement with CC2 energies for both the $L_a$ and $L_b$ excitations. Moreover, all of the LC-TDDFT methods preserve the correct ordering of electronic states between $n = 2$ and $n = 3$ benzene rings. In the case of the CAM-B3LYP functional though (which only has 65% HF exchange at long range), there are still some systematic discrepancies for the $L_a$ excitations which are still somewhat underestimated. Although the energetic ordering of the $L_a$ and $L_b$ states is correctly predicted by CAM-B3LYP, the energy differences are almost negligible, with only a 0.05 eV difference between the $L_a$ and $L_b$ states of naphthalene (compared to a ~0.2 eV difference with the full-exchange LC functionals). These observations strongly indicate that a range-separated partitioning of exchange alone, without 100% asymptotic HF exchange, is not sufficient, and a full asymptotic contribution of exchange is essential for accurately describing both the $L_a$ and $L_b$ excitations. Turning now to the global hybrids, Figure 3a shows that the B3LYP functional severely underestimates excitation energies (i.e. overestimates absorption wavelengths) for the $L_a$ electronic state. The situation is somewhat improved upon using the RMSE-optimized $a_0 = 0.50$ value in B3LYP$_{opt}$; however, this procedure results in $L_b$ excitations which are now *overestimated* and $L_a$ excitations which are still quite underestimated. Most importantly, both B3LYP and B3LYP$_{opt}$ give an incorrect ordering of electronic states – the crossing between $L_a$ and $L_b$ curves occurs much too early in both functionals, and the electronic symmetries in naphthalene have the wrong order. In general, the accuracy in excitation energies and trends is significantly improved with the LC scheme, while conventional hybrids are unable to reproduce the qualitative behavior in excitations *even if the fraction of HF exchange is optimized*.

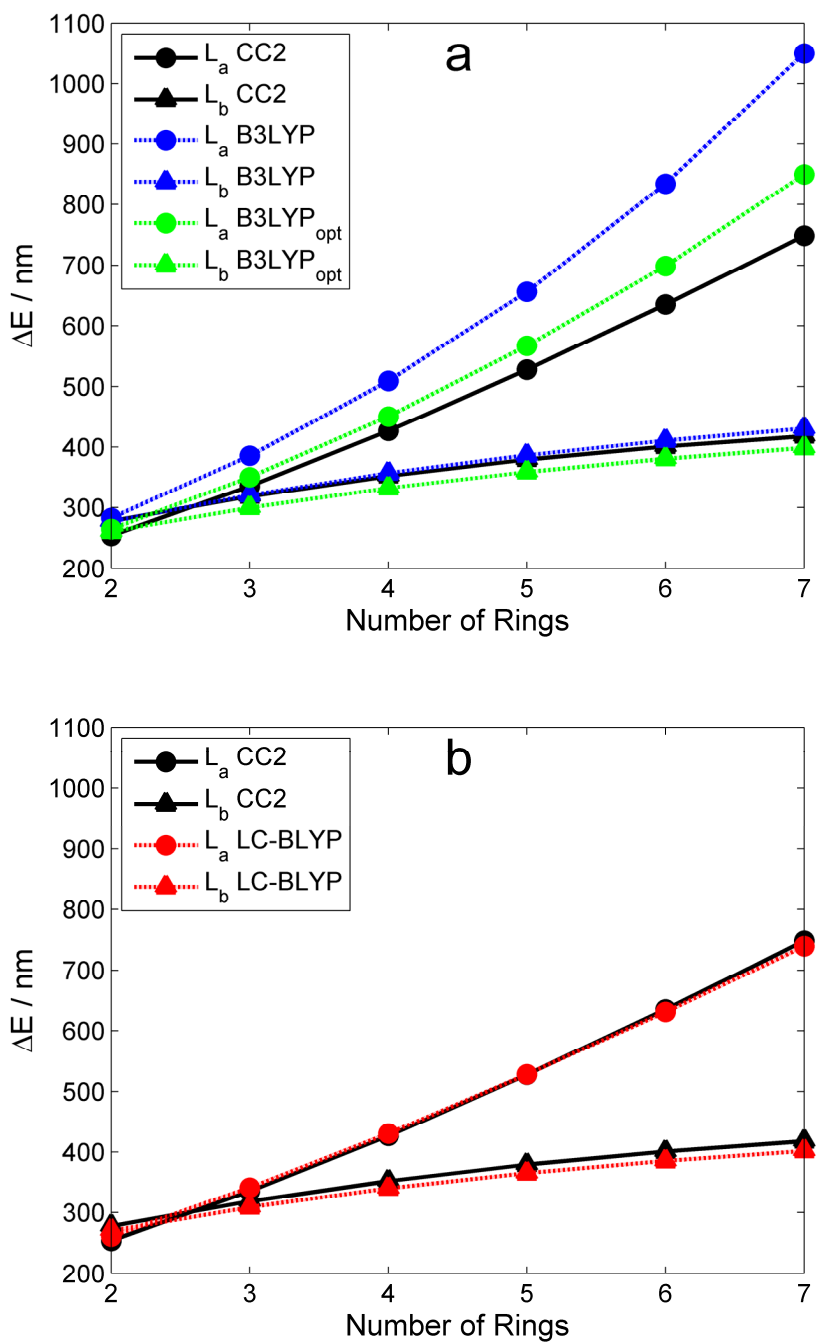

**Figure 3.** Comparison between TDDFT and CC2 excitation energies (in wavelength units) for (a) conventional global hybrid and (b) range-separated LC-BLYP functionals. The B3LYP$_{opt}$ functional denotes a modified B3LYP functional with a RMSE-optimized exchange fraction of $a_0 = 0.50$, as discussed in the main text.

From these results, it is interesting to note that long-range charge transfer is not responsible for the unexpected failure of B3LYP in these highly-symmetrical systems. In a recent benchmark study, Peach et al.[26] introduced a diagnostic test which quantifies the spatial overlap, $\Lambda$, between the occupied and virtual orbitals involved in an excitation. This diagnostic metric is typically used to post-process a converged TDDFT calculation, and has an intuitive form given by

$$\Lambda = \frac{\sum_{i,a}(X_{ia}+Y_{ia})^2 O_{ia}}{\sum_{i,a}(X_{ia}+Y_{ia})^2}. \tag{4}$$

In this expression, $X_{ia}$ and $Y_{ia}$ are the virtual-occupied and occupied-virtual transition amplitudes, respectively, and $O_{ia}$ is the spatial overlap integral of the moduli of the two orbitals, $O_{ia} = \int |\phi_i(\mathbf{r})||\phi_a(\mathbf{r})|d\mathbf{r}$. By construction, the diagnostic metric $\Lambda$ is bounded between 0 and 1, with small values signifying a long-range excitation and large values indicating a localized, short-range transition. Based on their extensive benchmarks, if $\Lambda$ is less than 0.3, indicating little overlap and significant long-range charge transfer character, hybrid functionals are predicted to yield inaccurate results. In Table 2, we computed the $\Lambda$ diagnostic for both the $L_a$ and $L_b$ states and found that all values were well above the 0.3 threshold (some of them even approaching 0.9), indicating a substantial overlap and no long-range charge transfer in these systems (it is rather interesting though that the diagnostic incorrectly predicts the $L_a$ excited state to be more accurately described than the $L_b$ state in both the B3LYP and LC-BLYP functionals). Thus, instead of long-range charge transfer from one end of the molecule to the other, we do find that the $L_a$ excitation involves a sizeable local rearrangement of electron density. In support of this assertion, Figure 4 depicts the electron density difference map ($\rho_{\text{excited}} - \rho_{\text{ground}}$) for the $L_a$ and $L_b$ excited states in pentacene computed at the CC2 level of theory (difference maps for the other acenes can be found in the Supporting Information). The electron density difference map gives a dynamic visualization of electronic rearrangement for a transition, with red regions (positively valued) denoting an accumulation of density and blue regions (negatively valued) representing a depletion of density upon excitation. As depicted in Figure 4, the $L_a$ state involves significantly more local charge redistribution than the higher-energy $L_b$ state. In contrast, the $L_b$ excited-state density is very similar to the ground state, as evidenced by the very small and sparsely-distributed isosurface regions. These CC2 difference densities confirm the long-held valence-bond viewpoint[70-72] that the $L_a$ state possesses an "ionic" character whereas the $L_b$ transition is primarily covalent in nature. Notice also that the length scale of charge redistribution is on the order of the carbon-carbon bond length (~1.4 Å), which is comparable to the length scale at which LC-BLYP predicts long-range HF exchange to dominate short-range DFT correlation ($1/\mu$ ~ 1.8 Å). Even though none of these transitions have long-range charge transfer character, our findings do support the physical interpretation that a range-separated contribution of full HF exchange on the length scale of the molecule is still necessary for accurately describing these local charge rearrangements.

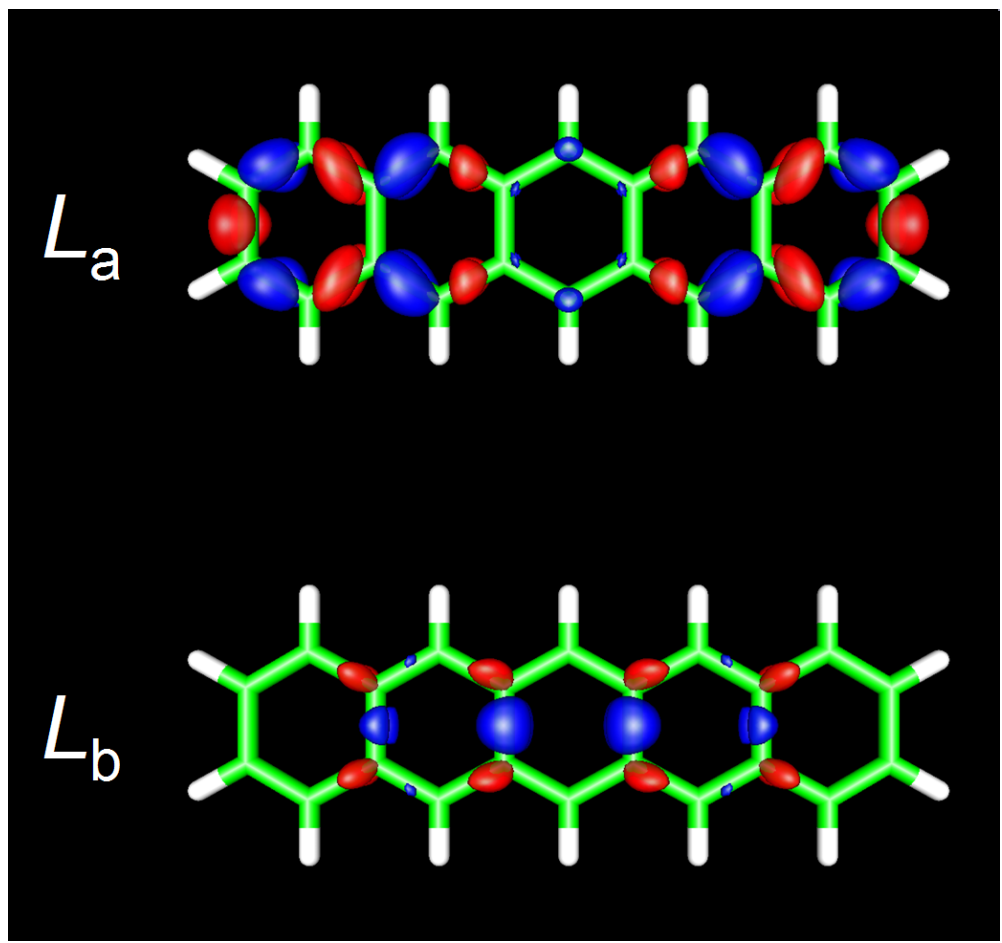

**Figure 4.** Electron density difference maps ($\rho_{\text{excited}} - \rho_{\text{ground}}$) for the $L_a$ and $L_b$ excited states of anthracene computed at the CC2 level of theory. Red regions denote a positive density difference (accumulation of density upon electronic excitation), and blue regions represent a negative density difference (depletion of density upon excitation). Both densities are plotted using the same isosurface contour value.

In order to provide further insight into these optoelectronic trends, we carried out an investigation of excitonic effects by analyzing electron-hole transition density matrices for the various excitations and TDDFT methods. Following the two-dimensional real-space analysis approach of Tretiak,[47-50] one can construct coordinate $\mathbf{Q_v}$ and momentum $\mathbf{P_v}$ matrices with elements given by

$$(Q_v)_{mn} = \langle \psi_v | c_m^\dagger c_n | \psi_g \rangle + \langle \psi_g | c_m^\dagger c_n | \psi_v \rangle \quad (5)$$

$$(P_v)_{mn} = \langle \psi_v | c_m^\dagger c_n | \psi_g \rangle - \langle \psi_g | c_m^\dagger c_n | \psi_v \rangle \quad (6)$$

where $\psi_g$ and $\psi_v$ are ground and excited states, respectively. The Fermi operators $c_i^\dagger$ and $c_i$ represent the creation and annihilation of an electron in the $i$th basis set orbital in $\psi$. For the acene systems analyzed in this work, the $\mathbf{Q_v}$ and $\mathbf{P_v}$ matrices each form a two-dimensional $xy$ grid over all the carbon sites along the $x$ and $y$ axes. The specific ordering of the carbon sites used in this work is shown in Figure 1. The $(Q_v)_{mn}$ coordinate matrix gives a measure of exciton delocalization between sites $m$ and $n$, and the $(P_v)_{mn}$ momentum matrix represents the probability amplitude of an electron-hole pair oscillation between carbon sites $m$ and $n$, respectively. Each of these matrices provides a complementary view of exciton delocalization and electron-hole coherence for optical transitions within the acene systems.

Figure 5 displays the absolute value of the coordinate density matrix elements, $|(Q_v)_{mn}|$, for the $L_a$ and $L_b$ excitation energies computed at the LC-BLYP level of theory. The $x$ and $y$ axes in this figure represent the benzene repeat units in the molecule, and the individual matrix elements are depicted by the various colors. Based on its construction, off-diagonal elements with large intensities represent widely-separated electron-hole pairs between different atoms. As shown in Figure 5, the $L_a$ density matrix has more off-diagonal elements than the corresponding $L_b$ excitation, whose matrix elements are primarily confined along the diagonal. These figures reflect the more delocalized nature of the $L_a$ state, in agreement with the electron density difference maps discussed previously. It is also important to note that all the transition density plots are symmetric along the counterdiagonal ($\searrow$), verifying that no long-range charge transfer occurs in these systems (an asymmetric transition density along the counterdiagonal implies more electrons than holes are localized on one side of the molecule).

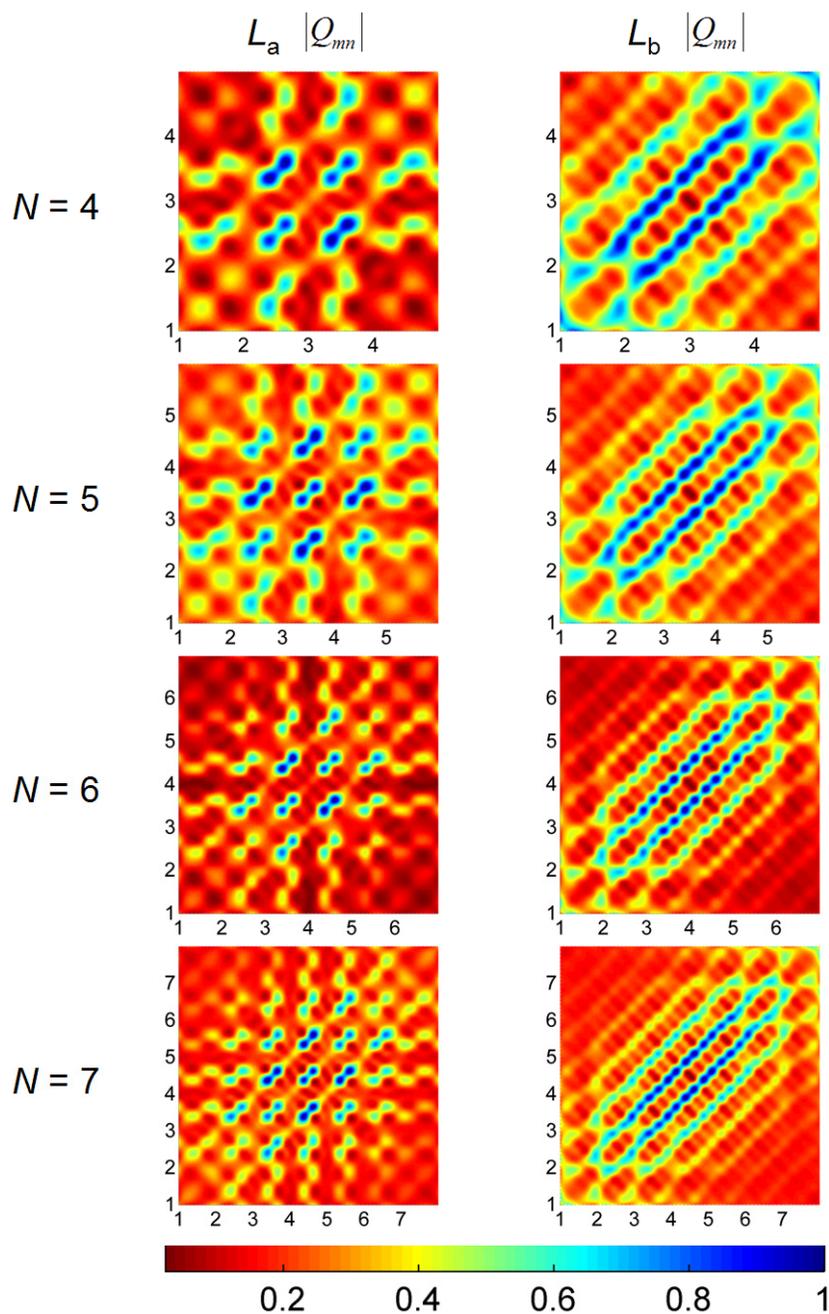

**Figure 5.** Contour plots of coordinate density matrices (**Q**) for the $L_a$ and $L_b$ excited states computed at the LC-BLYP level of theory. The x- and y-axis labels represent the number of benzene repeat units in the molecule. The elements of the coordinate matrix, $Q_{mn}$, give a measure of exciton delocalization between sites $m$ (x axis) and $n$ (y axis). The color scale is given at the bottom.

The coherence size, which characterizes the distance between an electron and a hole, is given by the width of the momentum density matrix, **P**$_v$. To compare excitonic effects between global and range-separated hybrids, we plot the absolute value of the momentum density matrix elements, $|(P_v)_{mn}|$, for both the B3LYP and LC-BLYP functionals in Figure 6 (transition density plots for all of the different functionals and excited states can be found in the Supporting Information). These figures show that the B3LYP functional gives a more delocalized density-matrix pattern and a larger coherence size compared to the LC-BLYP functional. Furthermore, the coherence size as predicted by the B3LYP functional is larger by nearly one repeat unit in comparison to the LC-BLYP results. These findings are consistent with the B3LYP formalism which only incorporates a global fraction of 20% HF exchange and, therefore, exhibits a $-0.2/r$ dependence for the exchange potential. As a result, the asymptotically-incorrect B3LYP exchange potential is not attractive enough, leading to an over-delocalized electron-hole pair and, therefore, an overestimated coherence size in the acene systems.

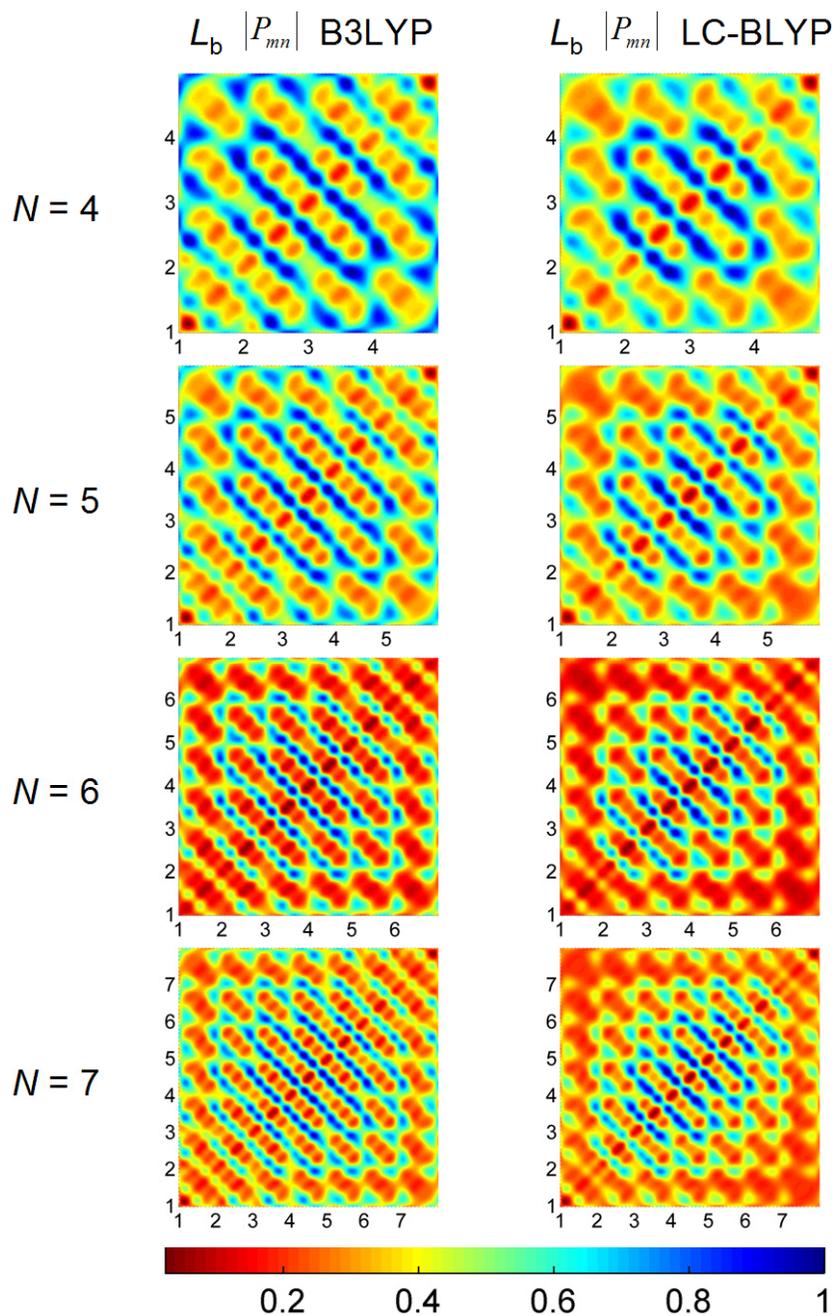

**Figure 6.** Contour plots of momentum density matrices (**P**) for the $L_b$ excited state computed at the B3LYP and LC-BLYP levels of theory. The *x*- and *y*-axis labels represent the number of benzene repeat units in the molecule. The elements of the momentum matrix, $P_{mn}$, represents the probability amplitude of an electron-hole pair oscillation between sites *m* (*x* axis) and *n* (*y* axis). The color scale is given at the bottom.

Finally, it is interesting to compare quasiparticle energy gaps predicted by both global hybrid and range-separated functionals in the acene systems. Within Kohn-Sham theory,[73] the quasiparticle gap can be approximated by the difference between the lowest unoccupied and highest unoccupied molecular orbital energies, $E_{LUMO} - E_{HOMO}$. Table 3 compares $-E_{LUMO}$, $-E_{HOMO}$, and the experimental ionization energies (IE) for the linear acenes computed at the B3LYP, CAM-B3LYP, and LC-BLYP levels of theory. From Kohn-Sham theory, it is well-known that an "exact functional" (if one had access to such a functional), would yield an ionization energy exactly equal to $-E_{HOMO}$. For the pentacene molecule as a specific example, the B3LYP functional provides a $-E_{HOMO}$ value of 4.78 eV which significantly underestimates the experimental ionization energy[74] of 6.61 eV. The $-E_{HOMO}$ values predicted by CAM-B3LYP are an improvement over the B3LYP energies, but the average deviation of -0.70 eV from the experimental IEs is still quite large. In contrast, the LC formalism, which incorporates a correct asymptotic behavior of the exchange potential by construction, gives $-E_{HOMO}$ values in exceptional agreement with all the experimental IEs, resulting in an impressive average deviation of 0.07 eV. These results complement our previous discussion of $L_a$ and $L_b$ excitation energies by further demonstrating that a full 100% asymptotic contribution of HF exchange is necessary to provide a consistent description of electronic properties in these systems. Furthermore, these findings demonstrate that the range-separated formalism with full asymptotic HF exchange is very self-consistent – *both* the excitation energies and quasiparticle properties in these systems are predicted accurately while simultaneously satisfying the energy constraints as required by Kohn-Sham theory.

**4. Conclusion**

In conclusion, the present study clearly indicates that both a range-separated partitioning as well as an asymptotically-correct contribution of exchange play a vital role in predicting optoelectronic properties in the linear acenes. Even though none of the excitations involve extended long-range charge transfer, we find that a range-separated contribution of full exchange is still necessary to accurately describe both the valence excitation energies and the $L_a \rightarrow L_b$ curve crossing in these simple systems. The results of our observations also strongly indicate that a range-separated partitioning of exchange by itself, without 100% asymptotic HF exchange (i.e., CAM-B3LYP), is not sufficient to accurately describe both the $L_a$ and $L_b$ state. Conversely, re-optimization of functional parameters towards 100% full exchange without range-separation in a global hybrid does not improve the situation either; in fact, this re-parameterization results in a corruption between exchange and correlation errors with trends in $L_a$ and $L_b$ excitations being even more poorly described. In particular, we find that global hybrid functionals overdelocalize excitons, underestimate quasiparticle energies, and are unable to reproduce general trends in both $L_a$ and $L_b$, even if the fraction of HF exchange is optimized. The most important results of our observations indicate that a simultaneous use of range-separated partitioning as well as a full contribution of exchange at large interelectronic distances is essential for accurately describing both the $L_a$ and $L_b$ states in these systems.

As acenes form the basis of nanoribbons and other polycyclic aromatic hydrocarbons,[75] this study serves an important role in determining which TDDFT methods are most appropriate for these systems, especially since wavefunction-based calculations on carbon nanostructures are still prohibitively demanding. Looking forward, it would be extremely interesting to see if the range-separated formalism also provides a similar accuracy for describing triplet states in acenes and other chromophores. While this study focused on only singlet excitations, further work is still needed to understand triplet excitations since exciton fission to low-lying triplet states ultimately control the electronic efficiencies in photovoltaic systems.[76] We are currently investigating these triplet states, with further calculations on extended organic light-harvesting systems,[9] to help predict the efficiencies of these materials. With this in mind, we anticipate that the LC-TDDFT technique will play a significant role in understanding and accurately predicting the optoelectronic properties in these novel nanostructures.

**Acknowledgment.** This research was supported in part by the National Science Foundation through TeraGrid resources (Grant No. TG-CHE1000066N) provided by the National Center for Supercomputing Applications. Funding for this effort was provided by the Laboratory Directed Research and Development (LDRD) program at Sandia National Laboratories, a multiprogram laboratory operated by Sandia Corporation, a Lockheed Martin Company, for the United States Department of Energy under contract DE-AC04-94AL85000.

**Supporting Information Available:** Electron density difference maps; contour plots of coordinate (Q) and momentum (P) density matrices; Cartesian coordinates of all the optimized structures. This material is available free of charge via the Internet at http://pubs.acs.org.

TABLE 1: Comparison of TDDFT, CC2, and experimental excitation energies [eV] (wavelengths [nm] are in parentheses) for the $L_a$ and $L_b$ states in the linear acenes. The mean absolute errors (MAE) relative to solvent-corrected experimental results are listed below each of the various methods. Excitation energies were computed with the cc-pVTZ basis with the same reference geometry for all of the different methods.

| Number of rings | B3LYP ($a_0 = 0.20$) | B3LYP$_{opt}$ ($a_0 = 0.50$) | CAM-B3LYP ($\alpha + \beta = 0.65$) | LC-BOP ($\mu = 0.29$) | LC-PBE ($\mu = 0.29$) | LC-$\omega$PBE ($\mu = 0.29$) | LC-BLYP ($\mu = 0.29$) | CC2 | Experiment[20] |
|---|---|---|---|---|---|---|---|---|---|
| $L_a$ State | | | | | | | | | |
| 2 | 4.39 (282) | 4.69 (264) | 4.68 (265) | 4.76 (260) | 5.05 (246) | 4.80 (258) | 4.76 (260) | 4.89 (254) | 4.66 (266) |
| 3 | 3.22 (385) | 3.54 (350) | 3.54 (350) | 3.64 (341) | 3.66 (339) | 3.67 (338) | 3.63 (342) | 3.70 (335) | 3.60 (344) |
| 4 | 2.44 (508) | 2.75 (451) | 2.77 (448) | 2.89 (429) | 2.89 (429) | 2.91 (426) | 2.88 (431) | 2.90 (428) | 2.88 (431) |
| 5 | 1.89 (656) | 2.19 (566) | 2.22 (558) | 2.36 (525) | 2.36 (525) | 2.38 (521) | 2.35 (528) | 2.35 (528) | 2.37 (523) |
| 6 | 1.49 (832) | 1.77 (700) | 1.83 (438) | 1.97 (629) | 1.98 (626) | 1.99 (623) | 1.96 (633) | 1.95 (636) | 2.02 (614) |
| 7 | 1.18 (1051) | 1.46 (849) | 1.53 (810) | 1.68 (738) | 1.69 (734) | 1.71 (725) | 1.68 (738) | 1.66 (747) | — |
| MAE (eV) | 0.42 | 0.13 | 0.11 | 0.04 | 0.10 | 0.06 | 0.04 | 0.09 | — |
| $L_b$ State | | | | | | | | | |
| 2 | 4.48 (277) | 4.75 (261) | 4.63 (268) | 4.59 (270) | 4.62 (268) | 4.61 (269) | 4.59 (270) | 4.47 (277) | 4.13 (300) |
| 3 | 3.87 (320) | 4.14 (299) | 4.04 (307) | 4.02 (308) | 4.04 (307) | 4.03 (308) | 4.02 (308) | 3.90 (318) | 3.64 (341) |
| 4 | 3.48 (357) | 3.73 (332) | 3.66 (339) | 3.65 (340) | 3.67 (338) | 3.66 (339) | 3.65 (340) | 3.52 (352) | 3.39 (366) |
| 5 | 3.21 (386) | 3.46 (358) | 3.40 (365) | 3.40 (365) | 3.41 (364) | 3.41 (364) | 3.39 (366) | 3.27 (379) | 3.12 (397) |
| 6 | 3.02 (411) | 3.26 (380) | 3.21 (386) | 3.22 (385) | 3.23 (384) | 3.23 (384) | 3.22 (385) | 3.09 (401) | 2.87 (432) |
| 7 | 2.88 (431) | 3.11 (399) | 3.08 (403) | 3.09 (401) | 3.10 (400) | 3.10 (400) | 3.08 (403) | 2.97 (417) | — |
| MAE (eV) | 0.18 | 0.44 | 0.36 | 0.35 | 0.36 | 0.36 | 0.34 | 0.22 | — |

**TABLE 2**: Comparison of the TDDFT $\Lambda$-overlap diagnostic for the $L_a$ and $L_b$ excited states in the linear acenes.

| Number of rings | B3LYP ($a_0 = 0.20$) | LC-BLYP ($\mu = 0.29$) |
|:---:|:---:|:---:|
| | $\Lambda$-overlap for $L_a$ states | |
| 2 | 0.89 | 0.89 |
| 3 | 0.88 | 0.88 |
| 4 | 0.88 | 0.88 |
| 5 | 0.89 | 0.89 |
| 6 | 0.89 | 0.89 |
| 7 | 0.90 | 0.90 |
| | $\Lambda$-overlap for $L_b$ states | |
| 2 | 0.65 | 0.64 |
| 3 | 0.65 | 0.65 |
| 4 | 0.63 | 0.63 |
| 5 | 0.62 | 0.62 |
| 6 | 0.60 | 0.61 |
| 7 | 0.59 | 0.60 |

**TABLE 3**: Comparison of $-E_{\text{LUMO}}$, $-E_{\text{HOMO}}$, and experimental ionization energies (IE) for the linear acenes computed at the B3LYP, CAM-B3LYP, and LC-BLYP levels of theory. The average deviation of $-E_{\text{HOMO}}$ relative to the experimental IE is listed below each of the various methods.

| | B3LYP ($a_0 = 0.20$) | | CAM-B3LYP ($\alpha + \beta = 0.65$) | | LC-BLYP ($\mu = 0.29$) | | Exp.[74] IE (eV) |
|---|---|---|---|---|---|---|---|
| Number of rings | $-E_{\text{LUMO}}$ (eV) | $-E_{\text{HOMO}}$ (eV) | $-E_{\text{LUMO}}$ (eV) | $-E_{\text{HOMO}}$ (eV) | $-E_{\text{LUMO}}$ (eV) | $-E_{\text{HOMO}}$ (eV) | |
| 2 | 1.21 | 6.00 | 0.10 | 7.40 | -0.60 | 8.21 | 8.14 |
| 3 | 1.85 | 5.43 | 0.84 | 6.72 | 0.17 | 7.51 | 7.44 |
| 4 | 2.29 | 5.05 | 1.34 | 6.27 | 0.69 | 7.03 | 6.97 |
| 5 | 2.59 | 4.78 | 1.71 | 5.95 | 1.07 | 6.69 | 6.63 |
| 6 | 2.81 | 4.59 | 1.98 | 5.71 | 1.36 | 6.44 | 6.36 |
| 7 | 2.98 | 4.45 | 2.18 | 5.53 | 1.57 | 6.25 | — |
| $\langle -E_{\text{HOMO}} - IE \rangle$ | — | -1.94 | — | -0.70 | — | 0.07 | — |